\documentclass[final,5p,times,twocolumn]{elsarticle}

\usepackage{amssymb}
\usepackage{color}
\usepackage{amsmath}
\usepackage{multirow}
\usepackage{booktabs} 
\usepackage{bm}

\usepackage[pdfstartview=FitH, colorlinks,
linkcolor={red!60!black!},
anchorcolor={green!80!black!},
urlcolor={blue!80!black!},
citecolor={blue!50!black!}]{hyperref}
\usepackage[table]{xcolor}

\journal{Physics Letters B}

\begin{document}


\begin{frontmatter}



\title{An {\it ab-initio} Gamow shell model approach with a core}


\author[ad1]{B. S. Hu}
\author[ad1]{Q. Wu}
\author[ad1]{J. G. Li}
\author[ad1]{Y. Z. Ma}
\author[ad1,ad2,ad3]{Z. H. Sun}
\author[ad4,ad5]{N. Michel}
\author[ad1]{F. R. Xu\corref{cor1}}

\address[ad1]{School of Physics, and State Key Laboratory of Nuclear Physics and Technology, Peking University, Beijing 100871, China}
\address[ad2]{Department of Physics and Astronomy, University of Tennessee, Knoxville, Tennessee 37996, USA}
\address[ad3]{Physics Division, Oak Ridge National Laboratory, Oak Ridge, Tennessee 37831, USA}
\address[ad4]{Institute of Modern Physics, Chinese Academy of Sciences, Lanzhou 730000, China}
\address[ad5]{School of Nuclear Science and Technology, University of Chinese Academy of Sciences, Beijing 100049, China}
\cortext[cor1]{frxu@pku.edu.cn}

\begin{abstract}

Gamow shell model (GSM) is usually performed within the Woods-Saxon (WS) basis in which the WS parameters need to be determined by fitting experimental single-particle energies including their resonance widths. In the multi-shell case, such a fit is difficult due to the lack of experimental data of cross-shell single-particle energies and widths. In this paper, we develop an {\it ab-initio} GSM by introducing the Gamow Hartree-Fock (GHF) basis that is obtained using the same interaction as the one used in the construction of the shell-model Hamiltonian. GSM makes use of the complex-momentum Berggren representation, then including resonance and continuum components. Hence, GSM gives a good description of weakly bound and unbound nuclei.  Starting from chiral effective field theory and employing many-body perturbation theory (MBPT) (called nondegenerate $\hat Q$-box folded-diagram renormalization) in the GHF basis, a multi-shell Hamiltonian ({\it sd-pf} shells in this work) can be constructed. The single-particle energies and their resonance widths can also been obtained using MBPT. We investigated $^{23-28}$O and $^{23-31}$F isotopes, for which multi-shell calculations are necessary. Calculations show that continuum effects and the inclusion of the {\it pf} shell are important elements to understand the structure of nuclei close to and beyond driplines.

\end{abstract}


\end{frontmatter}



\section{Introduction}
\label{sec:level1}
One of the current challenges in nuclear physics is to understand the microscopic structures of exotic nuclei in dripline regions.
Dripline nuclei belong to the category of open quantum systems, hence coupling to the  continuum affects profoundly their structures \cite{OKOLOWICZ2003271,0954-3899-36-1-013101}.
Neutron-rich oxygen and fluorine isotopes provide an excellent laboratory for studying the exotic properties of weakly bound and unbound nuclei. 
The neutron dripline in oxygen isotopes is situated at $^{24}$O \cite{HOFFMAN200917,PhysRevLett.102.152501}. By adding only one more proton, the fluorine chain has been reached with six more neutrons after $^{25}$F with $^{31}$F being the last known bound fluorine isotope \cite{SAKURAI1999180,Lukyanov_2002}. 
This dramatic change is referred to as the ``oxygen anomaly'', and the location of the oxygen dripline has been explained by the impact of three-nucleon forces (3NF) \cite{PhysRevLett.105.032501}.
However, the fluorine dripline has not yet been properly understood. 
From the conventional magic numbers $Z$=8 and $N$=20, the $^{28}$O should be an anticipated doubly magic nucleus. But several experiments \cite{SAKURAI1999180,TARASOV199764,PhysRevC.95.041301} point to the unbound character of this nucleus. The current experiments are limited by measurement precisions, and it has not yet been possible to precisely determine the energy and width of $^{28}$O.

There have been many theoretical investigations on oxygen (and fluorine) isotopes, including {\it ab-initio} calculations with chiral two and three-body interactions, such as in-medium similarity renormalization group (IMSRG) \cite{PhysRevLett.110.242501,PhysRevLett.113.142501,PhysRevLett.118.032502}, self-consistent Green’s function (SCGF) \cite{PhysRevLett.111.062501,PhysRevC.92.014306}, coupled cluster (CC) \cite{PhysRevLett.108.242501,PhysRevLett.113.142502}, important-truncation no-core shell model (IT-NCSM) \cite{TICHAI2018448}. 
The search for excited states in exotic nuclei beyond $^{24}$O ($N$=16) is of growing interest.
A recent experiment improved the description of the ground state of the barely unbound $^{26}$O, and provided also a measurement of the first 2$^{+}_{1}$ state \cite{PhysRevLett.116.102503}.
The new data were compared against several different theoretical calculations \cite{PhysRevLett.116.102503}.
It was suggested that further developments for {\it ab-initio} approaches are required to include continuum effects.
The spectrum of $^{26}$F is another challenge for theoretical description.
It was claimed in Refs.~\cite{PhysRevLett.110.082502,PhysRevC.96.054305} that the shell-model effective proton-neutron interaction should be reduced from the analysis of the $J^{\pi}$=$1^{+}_{1}$ - $4^{+}_{1}$ multiplet energies in $^{26}$F. 
This emphasizes the necessity to include the continuum coupling in shell-model effective interactions.
In addition, $pf$-shell configurations were found to be important in describing the excited states of neutron-rich oxygen and fluorine isotopes \cite{Holt2013,PhysRevC.95.041301}. Thus, the many-body dynamics, the continuum coupling and the inclusion of the $pf$ shell are important elements in understanding the structures of neutron-rich oxygen and fluorine isotopes around the dripline. 

The Gamow shell model (GSM) \cite{PhysRevLett.89.042501,PhysRevLett.89.042502} is a powerful tool for the description of weakly bound and unbound nuclei where continuum coupling is essential. 
This method is an extension of the traditional shell model into the complex-momentum (complex-$k$) space. It makes the use of the Berggren ensemble \cite{BERGGREN1968265,LIOTTA19961}, which treats bound states, unbound resonant states and nonresonant continuum states on an equal footing.
GSM has been successfully applied to weakly bound nuclei with phenomenological \citep{PhysRevLett.89.042501,PhysRevLett.89.042502,PhysRevC.96.024308} and realistic \cite{PhysRevC.73.064307,PhysRevC.80.051301,SUN2017227} interactions.

Using realistic nuclear forces, the derivation of the effective Hamiltonian in a given model space is challenging because the Berggren basis contours must be discretized with many scattering states for each partial wave. Scattering states bear the same nondegenerate multi-shell feature as cross-shell harmonic oscillator (HO) basis. Indeed, current non-perturbative calculations, e.g., using the IMSRG, have not been able to decouple the Hamiltonian in nondegenerate multi-shell spaces \cite{PhysRevC.96.034324}.
The many-body perturbation theory (MBPT) based on the extended Kuo-Krenciglowa (EKK) method \cite{TAKAYANAGI201161} provides a framework to derive multi-shell interactions \cite{TAKAYANAGI201161,PhysRevC.89.024313,PhysRevC.95.021304,SUN2017227}. 
However, a complete computation of (i) higher-order corrections in MBPT, (ii) full microscopic derivation of the effective Hamiltonian that includes single-particle energies (SPEs) and two-body shell-model interactions and (iii) the multi-shell effective Hamiltonian with continuum couplings, such as the $sdpf$-shell interaction, is still lacking.

In our previous work \cite{SUN2017227}, a full $\hat{Q}$-box folded-diagram renormalization within the nondegenerate complex-$k$ Berggren basis was developed to construct the $sd$-shell valence-space two-body interaction based on the realistic CD-Bonn potential. 
We used a Woods-Saxon (WS) potential to generate the Berggren basis, and took the leading-order WS SPEs as the one-body part of the effective Hamiltonian.
The neutron-rich oxygen isotopes up to $^{26}$O were well described by the newly developed method \cite{SUN2017227}. 
In the present paper, we replace the WS basis by the Gamow Hartree-Fock (GHF) \cite{PhysRevC.99.061302,PhysRevC.73.064307,PhysRevC.88.044318} basis and extend MBPT to construct both SPEs and two-body matrix elements of the GSM Hamiltonian. 
Using this method, we can avoid introducing the WS parameters in the Hamiltonian.  
The MBPT calculations performed within the GHF basis can provide a convergence of the same quality as that of nonperturbative approach \cite{PhysRevC.85.061304,Wu2019}. 
In addition, many perturbation diagrams are canceled out within the GHF basis, but not within the WS basis \cite{PhysRevC.94.014303,Wu2019,CORAGGIO200543}.

In this paper, perturbative effective Hamiltonians for the Gamow shell model calculations are derived for the first time from the nondegenerate third-order MBPT framework. We hope that the present MBPT+GSM framework can explain long-standing questions in oxygen and fluorine isotopes. As an introductive calculation, consequently, we will concentrate on the spectra of $^{26}$O and $^{26}$F. 

\section{Theoretical framework \label{sec:1}}
The intrinsic Hamiltonian of the $A$-nucleon system used in this work reads
\begin{eqnarray}
H=\displaystyle\sum_{i=1}^{A} \left(1-\dfrac{1}{A}\right) \dfrac{{\bm{p}_{i}}^{2}}{2m} +
\displaystyle\sum_{i<j}^{A}   \left(V_{\text{NN},ij}-\dfrac{\bm{p}_{i}\cdot\bm{p}_{j} }{mA} \right),
\label{Hamiltonian}
\end{eqnarray}
where $\bm{p}_i$ is the nucleon momentum in the laboratory coordinate and $m$ is the mass of the nucleon. The nucleon-nucleon (NN) interaction $V_{\text{NN}}$ used is the optimized chiral interaction NNLO$_{\text{opt}}$ \cite{PhysRevLett.110.192502}.  A good description of nuclear structure, including binding energies, excitation spectra, dripline positions and the neutron matter equation of state can indeed be obtained with the NNLO$_{\text{opt}}$ potential without resorting to three-body forces \cite{PhysRevLett.110.192502}.
In the calculation within the Gamow-Berggren representation, the choice of the potential to generate the one-body Berggren basis is crucial, contrary to standard shell model, where the harmonic oscillator basis is used. A WS potential has been used for that purpose in several applications \cite{PhysRevLett.89.042501,PhysRevLett.89.042502,PhysRevLett.120.212502,SUN2017227}.
In the present paper, the Berggren GHF basis is generated by diagonalizing the one-body HF Hamiltonian in a plane wave basis following a deformed contour $L^{+}$ in the complex-$k$ plane (see Refs.\cite{PhysRevC.99.061302,PhysRevC.73.064307,PhysRevC.88.044318} for the detail of the GHF approach).

In the MBPT calculations, we separate the $A$-nucleon Hamiltonian (\ref{Hamiltonian}) into a zero-order part $H_{0}$ and a perturbative part $H_{1}$,
\begin{equation}
H=H_{0}+(H-H_{0})=H_{0}+H_{1}.
\end{equation}
As mentioned above, we take a one-body HF Hamiltonian for $H_{0}$. We can then perform $\hat{Q}$-box folded-diagram calculations using the GHF basis.  
The GHF (Berggren) basis contains bound, resonant and scattering states, and the basis states are certainly not degenerate. 
As a consequence, in this work, we use the nondegenerate EKK method \cite{TAKAYANAGI201161,PhysRevC.89.024313,PhysRevC.95.021304} to construct the effective Hamiltonian $H_{\text{eff}}$ by iterating the following equation,
\begin{eqnarray}
H^{(n)}_{\text{eff}}=PH_{0}P+\hat{Q}(E)+ \sum_{k=1}^{\infty}\frac{1}{k!}\frac{d^{k}\hat{Q}(E)}{dE^{k}} \{H_{\text{eff}}^{(n-1)}-E\}^{k},
\label{EKK}
\end{eqnarray}
where $k$ indicates the $k$-th derivative and $E$ is the starting energy. The $\hat{Q}$-box is defined as
\begin{eqnarray}
\hat{Q}(E)=PH_{1}P+PH_{1}Q\dfrac{1}{E-QHQ}QH_{1}P.
\end{eqnarray}
$P$ and $Q$ represent the model space and the excluded space, respectively, with $P+Q=1$. 
In practical applications, the $\hat{Q}$-box is calculated perturbatively in terms of a power series in $H_{1}$  \cite{HjorthJensen1995125,Coraggio2012,Wu2019}.
In a fully microscopic approach, the one-body part of the effective Hamiltonian, namely shell-model SPEs, should be constructed from the realistic force in a manner similar to the two-body part. This procedure is called $\hat{S}$-box folded-diagram renormalization \cite{CORAGGIO200543}, where the $\hat{S}$-box is by definition the one-body part of the $\hat{Q}$-box. 
We perform the $\hat{S}$-box and $\hat{Q}$-box folded-diagram renormalizations to produce the shell-model SPEs and two-body matrix elements, respectively.
This fully microscopic approach produces a Hamiltonian for the Gamow shell model calculations in a valence space without additional adjustable parameters.  

\begin{figure}
\centering
\setlength{\abovecaptionskip}{6pt}
\setlength{\belowcaptionskip}{6pt}
\includegraphics[scale=0.32]{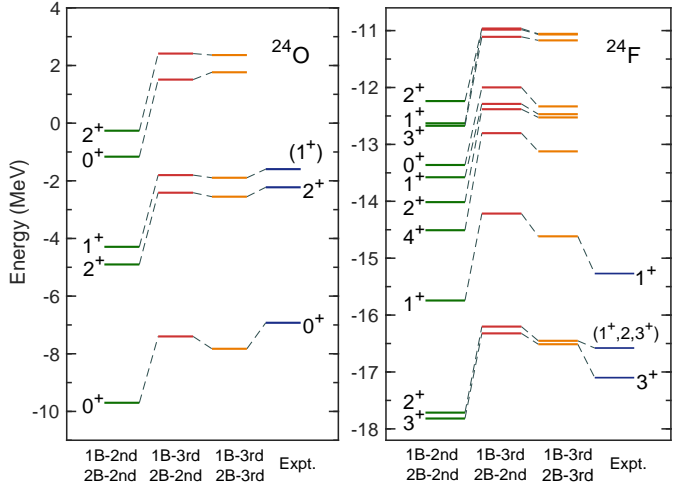}
\caption{Excitation energies of $^{24}$O and $^{24}$F calculated within the $sdpf_{7/2}$ valence-space Hamiltonian ($^{22}$O as the inert core) and compared with experimental data \cite{HOFFMAN200917,PhysRevC.83.031303,nndc}.  1B-2nd stands for the one-body part of valence-space Hamiltonian, constructed with MBPT by including up to second-order diagrams in $\hat{S}$-box; 2B-3rd  stands for the two-body part of valence-space Hamiltonian, constructed with MBPT by including up to third-order diagrams in $\hat{Q}$-box, and so on.}
\label{E_O24_F24_compare}
\end{figure}
Due to the large number of nonresonant continuum single-particle states, the full inclusion of third-order two-body $\hat{Q}$ diagrams is computationally very costly and often renders MBPT calculations impossible. 
In contrast, using the traditional HF basis expressed as a linear combination of HO basis wavefunctions, the number of orbitals is not so large, so that the third-order $\hat{Q}$ diagrams can be conveniently handled. We estimate the effect of the third-order two-body $\hat{Q}$ diagrams in the traditional HF basis, as shown in Figure~\ref{E_O24_F24_compare}. We can see that the uses of $\hat{S}$-box up to the third order and $\hat{Q}$-box up to the second order offer good descriptions of $^{24}$O and $^{24}$F. As a consequence, in the following calculations, we will use the one-body $\hat{S}$-box up to the third order and the two-body $\hat{Q}$-box up to the second order to construct the valence-space Hamiltonians within the GHF (Berggren) basis.

Let us give a summary of the calculation procedure: (i) we start from the chiral NN interaction NNLO$_{\text{opt}}$ \cite{PhysRevLett.110.192502} expressed in 13 major HO shells with the commonly used oscillator frequency $\hbar\Omega=20-24$ MeV \cite{PhysRevC.99.061302,HENDERSON2018468,Jiang:2019zkg}; (ii) we perform GHF calculations using the NNLO$_{\text{opt}}$ potential to produce the Berggren single-particle basis. The neutron $\nu d_{3/2}$, $\nu p_{1/2}$, $\nu p_{3/2}$ and $\nu f_{7/2}$ partial waves are treated in the GHF (Berggren) basis, while other neutron channels and all proton channels are handled in the traditional HF basis (i.e.,~obtained by diagonalizing the HF potential in a HO basis instead of in the Berggren basis). For the $\nu d_{3/2}$, $\nu p_{1/2}$, $\nu p_{3/2}$ channels including the resonances, we use the same contour $L^{+}$ discretized with 22 mesh points, as in our previous work \cite{SUN2017227}. We use 16 discretized states on the real-$k$ axis for the $f_{7/2}$ scattering channel. The convergence had been tested \cite{SUN2017227}; (iii) we demand $^{22}$O to be the inert core in our model. A full $\hat{S}$-box and $\hat{Q}$-box folded-diagram renormalization is performed in a nondegenerate complex-$k$ space to construct the effective Hamiltonian in the following model space: \{ (bound: $\pi 1s_{1/2}$, $\pi 0d_{3/2}$, $\pi 0d_{5/2}$, $\nu 1s_{1/2}$),  (resonance: $\nu 0d_{3/2}$, $\nu 1p_{1/2}$, $\nu 1p_{3/2}$) and (scattering continua: $\nu d_{3/2}$, $\nu p_{1/2}$, $\nu p_{3/2}$, $\nu f_{7/2}$) \}.
In the EKK calculation, the starting energy $E$ in Eq.~(\ref{EKK}) is taken equal to $-$20 MeV to avoid the poles of $\hat{Q}(E)$. We have verified that the calculations are not sensitive to the starting energy. For example, taking the starting energy $E$ from $-$15 to $-$25 MeV, the calculations of $^{26}$O ground and excited states change by about 0.05 MeV for energies and about 0.006 MeV for resonance widths; and (iv) we use the Jacobi-Davidson method \cite{DAVIDSON197587} to diagonalize the complex-symmetric GSM Hamiltonian.

For the SM calculation with a core, the SPEs of the SM space are critical to reproduce the experimental binding energies of nuclei. In a previous work \cite{Wu2019}, we show that the HF basis provides good convergences for many-body perturbation calculations, giving SPEs agreeing well with experimental data. Indeed, in the present calculations in the GHF basis with the GSM Hamiltonian, the $\hat{S}$-box perturbation up to third order can give reasonable SPEs compared with data, see Table~\ref{SPE}. A HF basis state can contain many different HO basis states (a superposition of many HO basis states), meaning that the HF basis includes more correlations. Therefore, it should be understood that the $\hat{S}$-box calculation in the HF (or GHF) basis can give reasonable SPEs. 

Although the Hamiltonian of Eq.(\ref{Hamiltonian}) is intrinsic, the SM wave function is written in laboratory coordinates. In a cross-shell SM space, the effect from the spurious center-of-mass (CoM) excitation should be considered. Moreover, the Berggren basis itself includes continuum channels, which belongs to a multi-shell problem. In the HO basis, the CoM correction can be done using the Lawson method making the use of the HO-CoM Hamiltonian $H_{\rm CoM}$ \cite{lawson}. It is, in principle, possible to generalize the Lawson method to the GSM. For this, one expands the one-body and two-body matrix elements of $H_{\rm CoM}$ in a HO basis, following the method described in Ref.\cite{PhysRevC.73.064307}. $H_{\rm CoM}$ can then be expressed in a Berggren basis representation and applied to GSM eigenvectors. It has been checked that this method provides an eigenstate free of CoM excitations if one expands many-body well-bound states within the Berggren basis, because the latter resemble the eigenstates obtained in the HO-basis SM. However, it cannot be generalized to halo and resonant many-body states, because the wave functions of halo and resonant states spread widely in space, which does not obey the characteristics of the HO wave functions. Indeed, as the Berggren basis space is not an $N\hbar \omega$ space, $H$ and $H_{\rm CoM}$ do not commute, so that the relative part of many-body wave functions can be affected by the use of the Lawson method for halo and resonant states. Consequently, it has been preferred not to apply the CoM correction in the present work. As a matter of fact, in our previous publications \cite{SUN2017227,PhysRevC.99.061302}, the calculations and discussions on the CoM correction have shown that the CoM effect within an intrinsic Hamiltonian is not significant for low-lying states. Therefore, we assume that the effects of CoM excitations can be neglected in the present calculations.

\begin{table}
\centering
\caption{
\label{SPE}
Single-particle energies (in MeV) given by the present GSM Hamiltonian, using the $\hat{S}$-box technique up to third order, compared with experimental data \cite{nndc,Wang_2017}. The resonance widths (in MeV) of resonant states are given in brackets.}
\begin{tabular}{lcc|lcc}
\hline \hline
\multicolumn{1}{c}{} & This work & Expt. & & This work & Expt. \\
\hline
$\nu 1s_{1/2}$ & $-$3.12 & $-$2.74 & $\pi 0d_{5/2}$ & $-$11.71 & $-$13.29  \\
\hline
$\nu 0d_{3/2}$ & 0.97 (0.11) & 1.26 (?) & $\pi 1s_{1/2}$ & $-$7.81 & $-$ \\
\hline
$\nu 1p_{3/2}$ & 2.60 (5.12) & $-$ & $\pi 0d_{3/2}$ & $-$7.17 & $-$ \\
\hline
$\nu 1p_{1/2}$ & 2.64 (5.21) & $-$ & & & \\
\hline \hline
\end{tabular}
\end{table}

\section{Calculations and discussions}

\begin{figure*}[t]
\setlength{\abovecaptionskip}{1pt}
\setlength{\belowcaptionskip}{1pt}
\hspace*{0.2cm}
\includegraphics[scale=0.48]{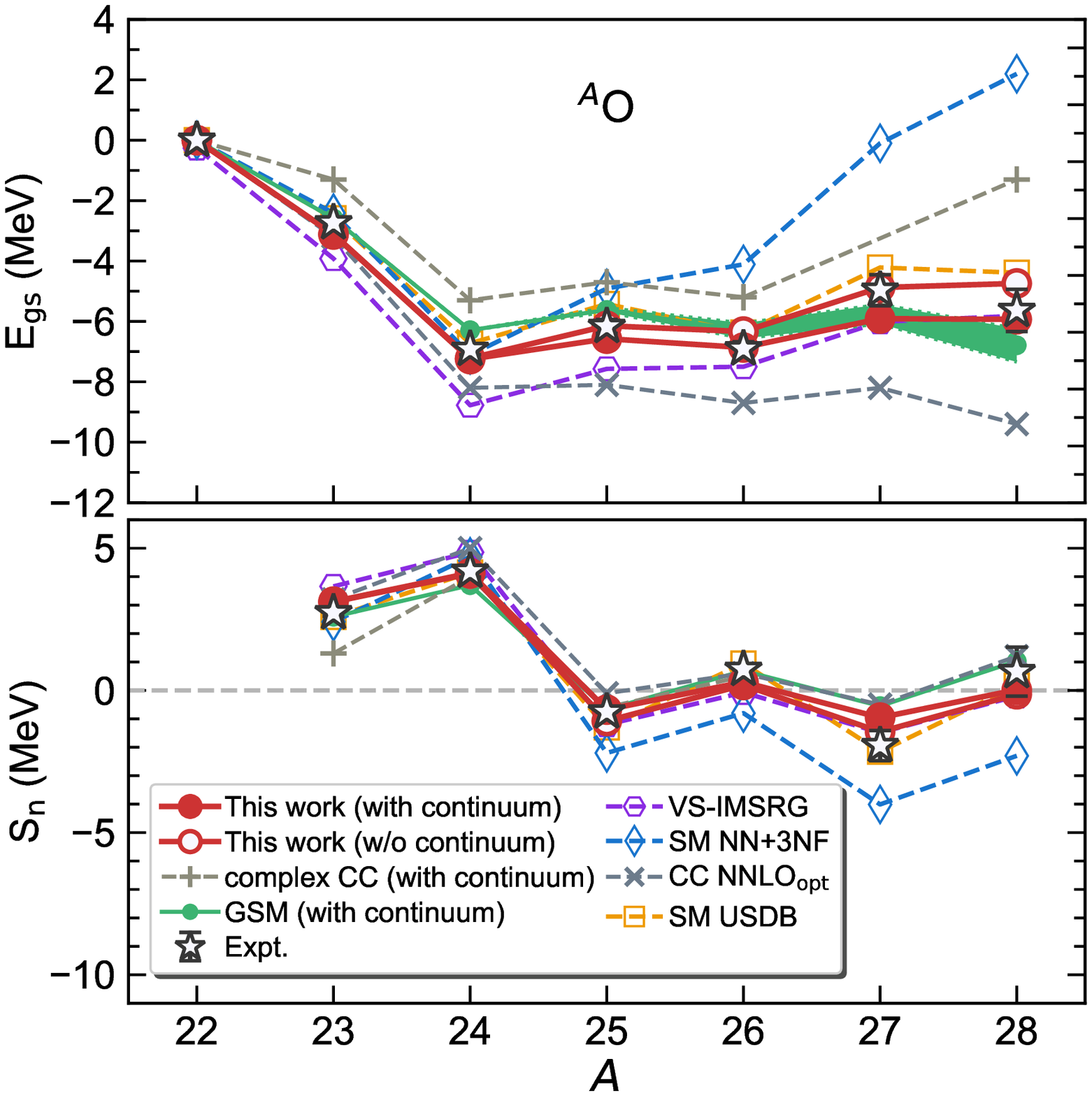}
\includegraphics[scale=0.48]{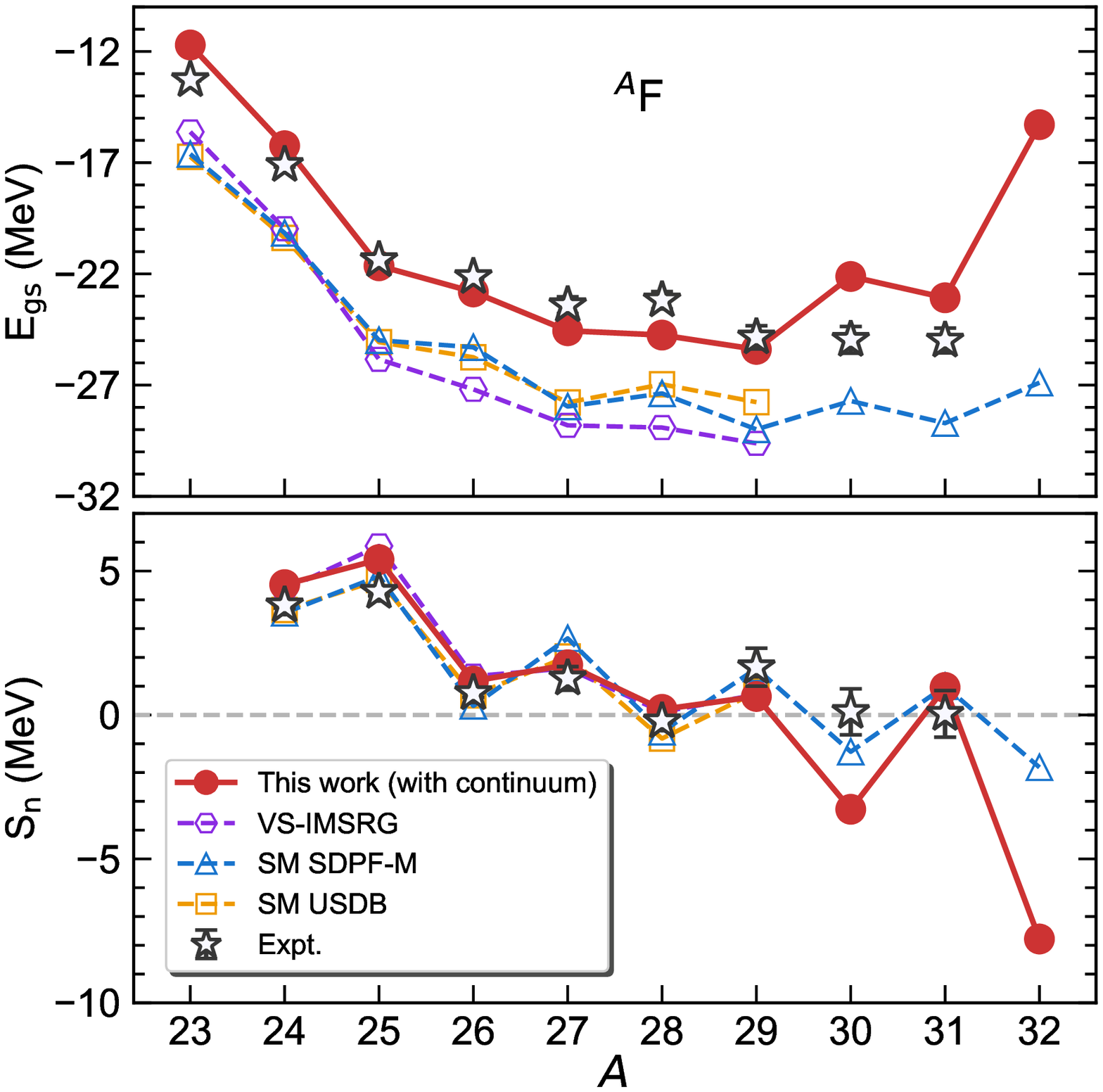}
\caption{Calculated ground-state energies (upper panel) with respect to the $^{22}$O core and one-neutron separation energies (lower panel) for oxygen and fluorine isotopes, compared with experimental data \cite{Wang_2017} (the AME2016 extrapolated values are taken for $^{27,28}$O and $^{30,31}$F) and other calculations: complex Coupled-Cluster (CC) with N$^3$LO(NN) and an effective 3NF \cite{PhysRevLett.108.242501}, CC with NNLO$_{\rm opt}$ interaction \cite{PhysRevLett.110.192502}, GSM \cite{PhysRevC.96.024308}, valence-space in-medium similarity renormalization group (VS-IMSRG) \cite{PhysRevC.93.051301}, SM with NN+3NF \cite{PhysRevLett.105.032501}, SM with USDB \cite{PhysRevC.74.034315} and SM with SDPF-M \cite{PhysRevC.60.054315}.}
\label{BE_Sn} 
\end{figure*}
Figure~\ref{BE_Sn} shows the calculated ground-state energies ($E_{\text{gs}}$) and single-neutron separation energies ($S_{\text{n}}$) for oxygen and fluorine isotopic chains, compared with experimental data and other theoretical calculations. 
As mentioned in the introduction, the location of the oxygen dripline has been reproduced in shell-model (SM) calculations \cite{PhysRevLett.105.032501}, where the 3NF acts repulsively and sets correctly the dripline at $^{24}$O (see SM with NN+3NF in Figure~\ref{BE_Sn}).
Indeed, all the theoretical calculations including 3NF effects properly give qualitatively the correct location of the oxygen neutron dripline. However, quantitative calculations can be significantly different by different models, especially for nuclei beyond the neutron dripline. One possible reason for the differences would be that open decay channels and particle continua are not included in calculations. For example, the resonance and continuum channels are not included in the SM with USDB interaction \cite{PhysRevC.74.034315}, in the SM with NN+3NF \cite{PhysRevLett.105.032501}, in the valence-space in-medium similarity renormalization group (VS-IMSRG) with NN+3NF \cite{PhysRevC.93.051301}. Weakly bound and unbound nuclei are prototypical open quantum systems in which the coupling to particle continuum is crucial and should be properly treated \cite{OKOLOWICZ2003271,0954-3899-36-1-013101}. 
The NNLO$_{\rm opt}$ potential was proposed and applied to oxygen isotopes using the CC approach in Ref.~\cite{PhysRevLett.110.192502}. We see that, with the same NNLO$_{\rm opt}$ potential, the present GSM and CC calculations ~\cite{PhysRevLett.110.192502} give consistent results.

The $\hat{Q}$-box folded-diagram renormalization does not provide interaction matrix elements for the energy calculation of the core. In the SM with a core, one obtains the energy of the valence particles, and usually the experimental energy of the core nucleus is taken if needed. The CC is a no-core calculation which can give the total energy of the nucleus. For the core nucleus $^{22}$O, the CC energy \cite{PhysRevLett.110.192502} is slightly higher than the datum. We estimate the energy of the core nucleus $^{22}$O by using the Rayleigh-Schr\"{o}dinger many-body perturbation theory (RSPT) which was employed in our previous work \cite{PhysRevC.94.014303}. Note that the existing RSPT works only for closed-shell nuclei and no continuum is included \cite{PhysRevC.94.014303}. With the NNLO$_{\rm opt}$ and resulted HF basis, the RSPT up to third-order correction gives a $^{22}$O ground-state energy of  $-$160.05 MeV which is in good agreement with the CC result of $\sim-$160 MeV using the same interaction NNLO$_{\rm opt}$\cite{PhysRevLett.110.192502}. However, in Fig.~\ref{BE_Sn} we see differences  between the SM (or GSM) and CC  energies with the same NNLO$_{\rm opt}$ for heavy oxygen isotopes. In fact, such differences have been seen in Ref.~\cite{PhysRevLett.113.142502} where the SM calculations with the coulped-cluster effective interaction (CCEI) give less bound energies than the CC calculations. The present results are similar to those in Ref.~\cite{PhysRevLett.113.142502}. It was commented \cite{PhysRevLett.113.142502} that the induced 3NF may play a role in many-body calculations within a limited model space where  an interaction renormalization (e.g., $\hat Q$-box folded diagram) needs to be carried out. Indeed, the shell-model coupled-cluster calculations with the inclusion of the induced 3NF can improve the agreement between SM and CC calculations \cite{PhysRevC.98.054320}. The SM can well treat many-nucleon  correlations in the valence space. The CC calculations were truncated with singles, doubles and leading order
three-particle–three-hole excitations, and the equation-of-motion approach was combined for open-shell nuclei \cite{PhysRevLett.110.192502}. Therefore, we may expect to see some differences in the SM and CC calculations.
In the complex CC calculation \cite{PhysRevLett.108.242501} which includes the continuum effect, the chiral N$^3$LO(NN) and a density-dependent effective 3NF were used.  

Our MBPT+GSM calculation gives the correct location of the neutron dripline of oxygen isotopes and also the best description of the unbound nuclei $^{25,26}$O which are beyond the neutron dripline (see the left panel of Figure~\ref{BE_Sn}). Two other calculations employing the Gamow-Berggren representation are also displayed in Figure~\ref{BE_Sn} for comparison. One is the complex coupled-cluster model (CC) \cite{PhysRevLett.108.242501} using a schematic 3NF, which reproduces the neutron dripline but deviates from experimental energies. The other is GSM \cite{PhysRevC.96.024308} using a phenomenological effective interaction fitted to the bound states and resonances of $^{23-26}$O, assuming an inert core of $^{22}$O. A good description of energies was obtained therein by the GSM, and it was predicted that the ground state of $^{28}$O is very weakly bound or slightly unbound. However, our calculation gives different predictions for the $^{27,28}$O ground states.

We predict from our calculations that $^{28}$O is weakly unbound compared to $^{26}$O and is more bound than $^{27}$O. The unbound nature of $^{25-28}$O is clearly seen from $S_{\text{n}}$ in Figure~\ref{BE_Sn}. In Ref.~\cite{PhysRevC.84.021303}, Grigorenko {\it et al.} addressed possible pure four-neutron radioactivity in $^{28}$O with $S_{\text{4n}}<0$ but $\left\{ S_{\text{n}},S_{\text{2n}},S_{\text{3n}} \right\}>0$. In our $^{28}$O calculation, we obtain $\left\{S_{\text{2n}}, S_{\text{3n}}, S_\text{{4n}}\right\}<0$ but $S_{\text{n}}>0$, therefore, 2n, 3n and 4n emissions are possible. We find that the main configuration in $^{28}$O ground state is $(\nu 1s_{1/2})^{2}(\nu 0d_{3/2})^{4}$.
Due to the unbound character of the $\nu 0d_{3/2}$ resonant single-particle state, the $^{28}$O ground state is unbound and its particle-emission width is about several tens of keV.  
To show the effects of the $\nu 0d_{3/2}$ decaying resonance and scattering continua, we did a comparison between the calculations done with and without continuum in Figure~\ref{BE_Sn}. It is seen that the continuum effect becomes larger with increasing neutron number. The continuum effect pulls the ground-state energy down by about 0.5 MeV in $^{26}$O and 1 MeV in $^{28}$O.

The right panels of Figure~\ref{BE_Sn} give the results of fluorine isotopes. Few {\it ab-initio} calculations with continuum coupling have been applied to the isotope chain. 
We give the first systematic calculations with continuum from first principles. 
For comparison, we also show the SM calculations with the commonly used effective interactions, USDB \cite{PhysRevC.74.034315} and SDPF-M \cite{PhysRevC.60.054315}. All ground-state energies in Figure~\ref{BE_Sn} are given relatively to the $^{22}$O binding energy. 
Due to the presence of the scattering continuum, the dimension of the GSM Hamiltonian matrix grows rapidly with the number of valence particles. 
For example, the dimension of the $^{29}$F ground state calculated in the complex-$k$ $sdpf_{7/2}$ valence space is $d \approx 1.95 \times 10^{10}$, which one cannot handle in GSM for the moment. Consequently, in practical calculations, we allow at most two valence particles in the continuum states from $^{29}$F to $^{32}$F. This truncation may be one of the reasons why our calculations cannot give a good description of binding energies after $^{29}$F. The experiments \cite{SAKURAI1999180,Lukyanov_2002} have revealed that the $^{31}$F lies at the neutron dripline. Although our results show that $^{31}$F is lower in energy than $^{30}$F, $^{31}$F is still unbound (against two-neutron emission) compared to $^{29}$F. However, our NNLO$_{\rm opt}$ calculations provide good descriptions of ground-state energies for $^{23-28}$F.
We have also calculated effective single-particle energies using the formula given in Ref.~\cite{YUAN2016237}. We find that $\nu 0d_{3/2}$ is unbound in the oxygen isotopes, while $\nu 0d_{3/2}$ becomes bound in fluorine isotopes with a single-particle energy of about $-1$ MeV due to the proton-neutron interaction.
This partially explains why the neutron dripline appears with about six more neutrons in fluorine isotopes than in oxygen isotopes.

\begin{figure}[h]
\centering
\setlength{\abovecaptionskip}{6pt}
\setlength{\belowcaptionskip}{6pt}
\includegraphics[scale=0.35]{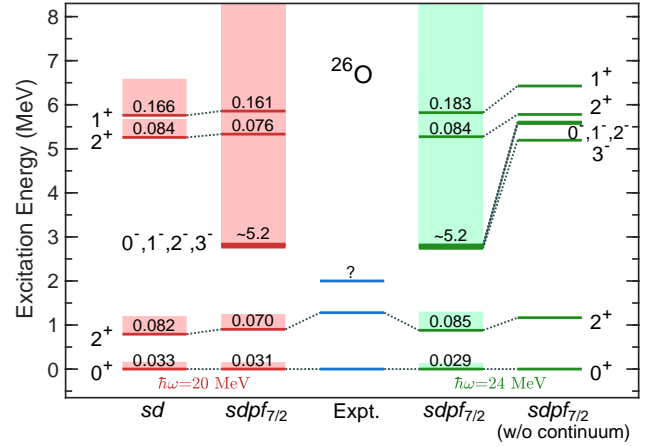}
\caption{The $^{26}$O spectrum calculated in different model spaces with the HO basis parameter $\hbar\omega$=20 MeV and $\hbar\omega$=24 MeV. On the figure, $sd$ indicates that the valence space is built from the bound $\nu 1s_{1/2}$, resonance $\nu 0d_{3/2}$ and neutron scattering continuum $d_{3/2}$ states, whereas $sdpf_{7/2}$ refers to a space generated by the bound $\nu 1s_{1/2}$, resonance $\nu 0d_{3/2}$, $\nu 1p_{1/2}$ and $\nu 1p_{3/2}$, and neutron continuum $d_{3/2}$, $p_{1/2}$, $p_{3/2}$, $f_{7/2}$ states. Unbound states are indicated by shading, with the numerical values of their widths (in MeV) explicitly written. Experimental data are taken from \cite{PhysRevLett.116.102503,nndc}.}
\label{Ex_O26}
\end{figure}
Let us now consider excited states. 
As mentioned in the introduction, the spectrum of $^{26}$O can provide a stringent test for the nuclear interaction and many-body correlations at extreme neutron-proton asymmetry. 
In order to clarify the various effects arising from the particle continuum and $pf$-shell configurations, we did calculations in different model spaces using the MBPT+GSM framework, as shown in Figure~\ref{Ex_O26}.
The left two columns compare the continuum-coupled result with and without $pf$-shell configuration mixing. The $pf$-shell configurations slightly push excited states in the upper energy region, but this effect is not very pronounced.
It is about 0.25 MeV for binding energy, which is smaller than the 0.5 MeV value stated in Ref.\cite{PhysRevC.96.024308}.
It is clear from the right two columns of Figure~\ref{Ex_O26} that continuum coupling pushes excited states down in the spectrum. Continuum coupling is larger for negative-parity states than for positive-parity states. This is because the negative-parity states are dominated by $p$ partial waves which have a small centrifugal barrier ($l$=1) and hence leads to a strong continuum coupling. Moreover, radial wave functions bearing one node increase the radial extension of many-body wave functions \cite{Hagen2016}.  
Finally, our calculations including both continuum and $pf$-shell configurations provide a ground state with about a 30 keV resonance width, and satisfactorily reproduce the first $2^{+}$ excited state as well.
Besides, we predict that there are superposed resonant states with $J^{\pi}=0^{-}, 1^{-}, 2^{-}, 3^{-}$ at energies $\sim2.8$ MeV and widths $\sim5.2$ MeV.

\begin{figure}[t]
\centering
\setlength{\abovecaptionskip}{6pt}
\setlength{\belowcaptionskip}{6pt}
\includegraphics[scale=0.35]{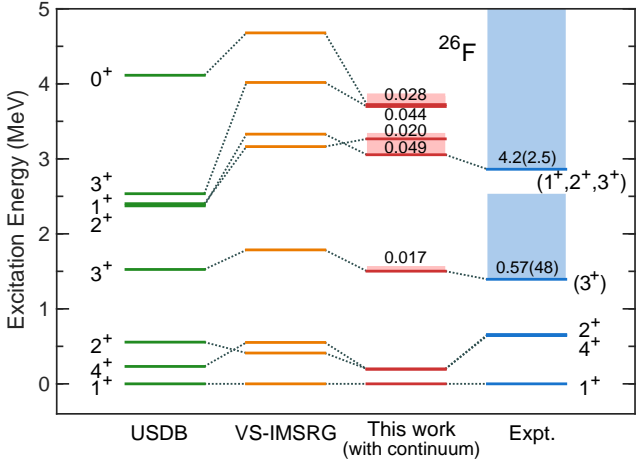}
\caption{Energy spectrum for $^{26}$F calculated with MPBT+GSM and compared with experimental data \cite{PhysRevLett.110.082502,PhysRevC.96.054305} and other calculations where continuum coupling is discarded: SM with USDB interaction and VS-IMSRG with NN+3NF.}
\label{Ex_F26}
\end{figure}
The spectrum of $^{26}$F is presented in Figure~\ref{Ex_F26} and compared to experimental data \cite{PhysRevLett.110.082502,PhysRevC.96.054305}.  
Considering the new shell closure discovered with $^{24}$O \cite{HOFFMAN200917,PhysRevLett.102.152501}, $^{26}$F can simply be treated as one valence proton and one valence neutron outside the $^{24}$O core. In particular, the $J^{\pi}=1^{+}_{1}-4^{+}_{1}$ multiple in the $^{26}$F spectrum is approximately generated by the coupling of a bound proton in the $\pi0d_{5/2}$ orbit and an unbound neutron in the $\nu0d_{3/2}$ orbit. From this simple two-particle description, we can directly extract the experimental and calculated residual interaction elements as $\langle\pi0d_{5/2},\nu0d_{3/2};J|V|\pi0d_{5/2},\nu0d_{3/2};J\rangle=[E_{J}(^{26}\text{F})-E_{0^{+}}(^{24}\text{O})]-[E_{5/2^{+}}(^{25}\text{F})-E_{0^{+}}(^{24}\text{O})]-[E_{3/2^{+}}(^{25}\text{O})-E_{0^{+}}(^{24}\text{O})]$. A good description of the multiplet provides a stringent test of the theoretical proton-neutron force.
Indeed, in our calculation, the $1^{+}_{1}-4^{+}_{1}$ multiplet is mainly generated by the $(\pi0d_{5/2})^{1}(\nu0d_{3/2})^{1}$ configuration. 
Compared to the SM calculations with the {\it ab-initio} VS-IMSRG and phenomenological USDB effective interactions,
our results can reproduce the isomeric state $4^{+}_{1}$ and give nearly degenerate $2^{+}_{1}$ and $4^{+}_{1}$ states, similarly to experimental data \cite{PhysRevLett.110.082502,PhysRevC.96.054305}. This points to that our effective Hamiltonian can generate a good proton-neutron interaction. 

Table~\ref{interaction} shows the calculated and experimental proton-neutron interaction matrix elements.
We see that the present calculations reproduce well the experimental data. 
The deduced amplitudes for the experimental residual interaction matrix elements are indeed smaller than those provided by calculations without the continuum coupling. However, theoretical effective interaction matrix elements depend to some extent on the model space chosen. The experimental states are more compressed as well. This points out the necessity to include the continuum coupling to weaken the matrix elements of the proton-neutron interaction.  
The comparison between calculations done with and without the continuum coupling in Table~\ref{interaction} clearly shows that continuum neutron-neutron correlations can affect the proton-neutron interaction. This renders our residual interaction matrix elements close to experimental data. 
The remaining differences between calculations and experimental data may be caused by the exclusion of high-lying configurations in the calculated $^{24}$O wave function. 
Indeed, in our calculations, we find that a tiny percentage of the configurations do not belong to the $^{24}$O+one-proton+one-neutron configurations. 
\begin{table}
\centering
\caption{
\label{interaction}
Calculated and experimental residual interaction elements (in MeV) in $^{26}$F.
The experimental data are deduced from Refs. \cite{PhysRevLett.110.082502,PhysRevC.96.054305,nndc}.
}
\begin{tabular}{ccccc p{cm}}
\hline \hline
\multicolumn{1}{c}{} & \multicolumn{4}{c}{$\langle\nu0d_{3/2},\pi0d_{5/2};J|V|\nu0d_{3/2},\pi0d_{5/2};J\rangle$} \\
\cmidrule{2-5}
\multicolumn{1}{c}{} &$J$=1 & $J$=2 & $J$=3 & $J$=4 \\
\cmidrule{1-5}
USDB \cite{PhysRevC.74.034315} & $-$1.98 & $-$1.43 & $-$0.46 & $-$1.75 \\
\cmidrule{1-5}
SDPF-M \cite{PhysRevC.60.054315} & $-$2.27 & $-$1.53 & $-$0.59 & $-$1.90 \\
\cmidrule{1-5}
VS-IMSRG \cite{PhysRevC.93.051301} & $-$2.67 & $-$2.26 & $-$0.88 & $-$2.11 \\
\cmidrule{1-5}
\shortstack{This work\\(w/o continuum)}  & $-$2.28 & $-$2.03 &$-$0.52 & $-$2.04 \\
\cmidrule{1-5}
\shortstack{This work\\(with continuum)} & $-$1.85 & $-$1.65 & $-$0.35 & $-$1.66 \\
\cmidrule{1-5}
Expt. & $-$1.85 & $-$1.19 & $-$0.45 & $-$1.21 \\
\hline \hline
\end{tabular}
\end{table}
 
\section{Summary}
We have presented the first {\it ab-initio} construction of a $sdpf$-multi-shell Hamiltonian in the Gamow shell model (GSM) framework based on the chiral NNLO$_{\text{opt}}$ interaction.
The Gamow Hartree-Fock (GHF) approach is performed to generate the Berggren basis, which treats bound, resonance and scattering states on an equal footing in the complex-momentum plane.
Within the nondegenerate GHF (Berggren) basis, the single-particle energies and two-body matrix elements of the GSM Hamiltonian are simultaneously derived from the many-body perturbation theory (MBPT), namely the $\hat{Q}$-box (or $\hat{S}$-box) full-folded-diagram method by the extended Kuo-Krenciglowa (EKK) method. We show that the GHF basis provides good convergence in MBPT calculations. Consequently, the MBPT+GSM framework used along with the GHF provides rigorous and tractable calculations for weakly bound and unbound open quantum systems.

The first systematic calculations for oxygen and fluorine isotopic chains arising from first principles including resonance and continuum states have been made. By taking a core of $^{22}$O, the unbound nature of $^{25-28}$O, lying beyond the neutron dripline, is nicely described. In particular, $^{28}$O is predicted to be weakly unbound by two-neutron emission, which can guide future experiments. 
The binding energies of $^{23-31}$F are also satisfactorily reproduced, even though the ground states of $^{29-31}$F are too high in energy.
By calculating the spectra of $^{26}$O and $^{26}$F, we could perform a stringent study of the $sdpf$-valence-space Hamiltonian compared to the case where $pf$ configurations or continuum effects are omitted. 
The locations of the neutron dripline and positions of excited states in the neutron-rich oxygen and fluorine isotopes can be determined by the proper inclusions of many-body correlations, $pf$-shell configurations and continuum degrees of freedom. 
However, the description of the neutron dripline for fluorine isotopes remains challenging in {\it ab-initio} calculations.

\section*{Acknowledgements}
This work has been supported by
the National Key R${\&}$D Program of China under Grant No. 2018YFA0404401;
the National Natural Science Foundation of China under Grants No. 11835001, No. 11921006, No. 11575007 and No. 11847203;
China Postdoctoral Science Foundation under Grant No. 2018M630018;
and the CUSTIPEN (China-U.S. Theory Institute for Physics with Exotic Nuclei) funded by the U.S.  Department of Energy,
Office of Science under Grant No. DE-SC0009971.
We acknowledge the High-performance Computing Platform of Peking University for providing computational resources.



  \bibliographystyle{elsarticle-num_noURL}
  \bibliography{references}





\end{document}